\documentclass[aps,prl,twocolumn,superscriptaddress,showpacs]{revtex4-1}

\pdfoutput=1
\usepackage {graphicx}
\usepackage{amsmath}
\usepackage{wrapfig}
\usepackage{epstopdf}
\usepackage{color}

\begin{document}


\title{Multiple barriers in forced rupture of protein complexes}


\author{Changbong Hyeon}
\affiliation{Korea Institute for Advanced Study, Seoul 130-722, Republic of Korea}
\author{D. Thirumalai}
\affiliation{Biophysics Program, Institute for Physical Science and Technology, University of Maryland, College Park, MD 20742, USA}


\begin{abstract}
Curvatures in the most probable rupture force ($f^*$) versus log-loading rate ($\log{r_f}$) observed in dynamic force spectroscopy (DFS) on biomolecular complexes are interpreted using a one-dimensional free energy profile with multiple barriers or a single barrier with force-dependent transition state. Here, we provide a criterion to select one scenario over another. If the rupture dynamics occurs by crossing a single barrier in a physical free energy profile describing unbinding, the exponent $\nu$, from $(1- f^*/f_c)^{1/\nu}\sim(\log r_f)$ with $f_c$ being a critical force in the absence of force, is restricted to $0.5 \leq \nu \leq 1$. For biotin-ligand complexes and leukocyte-associated antigen-1 bound to intercellular adhesion molecules, which display large curvature in the DFS data, fits to experimental data yield $\nu<0.5$, suggesting that ligand unbinding is associated with multiple-barrier crossing.
\end{abstract}


\maketitle


\section*{\large{Introduction}}

Single molecule pulling experiments have generated a wealth of data, which can be used to probe aspects of folding that were not previously possible \cite{Thirumalai10ARB,Borgia08ARBiochem,Fernandez09PNAS}. In addition, DFS has been used to decipher the energy landscape of molecular complexes by measuring the rupture force ($f$) by linearly increasing load at a rate $r_f$ (= $df/dt$). Because of the stochastic nature of the unbinding events, $f$ varies from one complex (or realization)  to another, giving rise to an $r_f$-dependent rupture force distribution ($P(f)$). For a molecular complex obeying Bell's formula, $k(f) = k_{off}\exp{(fx^{\ddagger}/k_BT)}$, 
Evans and Ritchie showed that the most probable force is $f^* = (k_BT/x^{\ddagger})\log (r_fx^{\ddagger}/k_{off}k_BT)$ \cite{Evans97BJ}, where $x^{\ddagger}(=x_{ts}-x_b)$ is the location of the transition state ($x_{ts}$) from the bound state ($x_b$) projected along the pulling coordinate and $k_{off}$ is the unbinding rate in the absence of force. However, Bell's formula is applicable only if the molecular complexes are mechanically brittle or if the applied tension is sufficiently small that $x^{\ddagger}$ does not shift upon application of force \cite{Hyeon07JP}. More generally, $f^*$ follows a $(\log r_f)^{\nu}$ dependence where $\nu$ depends on the details of the assumed one dimensional (1D) model potential \cite{Garg95PRB,HummerBJ03,Dudko03PNAS,Sheng05JCP,Dudko06PRL, Dudko07BJ,Lin07PRL,Friddle08PRL}. 
The basic assumption in all these works is that a single free energy barrier along the pulling coordinate is sufficient to describe force-driven rupture of the bound complex.

Sometime ago Merkel \emph{et al.} used DFS to probe load dependent strength of biotin bound to ligands, streptavidin and avidin \cite{EvansNature99}, showing that over six orders of variation in $r_f$ (from about $10^{-2}$ to $r_f$ in excess of $10^4$ pN/s) the plot of $f^*$ versus $\log r_f$ ($[f^*,\log r_f]$ plot) varies nonlinearly for both ligands. 
We note parenthetically that it is also common to observe curvature in unfolding rates of proteins 
when the $r_f$ is varied \cite{Schlierf06BJ}. By careful data analysis combined with molecular dynamics simulations they proposed an energy landscape for the complex, with multiple energy barriers \cite{EvansNature99}. A similar picture emerges in the rupture of intercellular adhesion molecules (ICAM-1 and ICAM-2) bound to leukocyte function-associated antigen-1 (LFA-1) upon application of force \cite{Moy06Biomacromolecules}.  

In principle, however, nonlinearity in $[f^*,\log r_f]$ plot could also arise from load dependent variation in $x^{\ddagger}$ \cite{Hyeon06BJ} in a 1D energy landscape with a single barrier \cite{Hyeon07JP,Garg95PRB,HummerBJ03,Dudko03PNAS,Sheng05JCP,Dudko06PRL,Lin07PRL,Friddle08PRL,Hyeon06BJ}. 
A theoretical model describing force-induced escape from a bound state with a single barrier in a cubic potential ($\nu=2/3$) has been used to rationalize the biotin-ligand data by identifying various linear regimes demarcated by $r_f$ \cite{Sheng05JCP}. However, in the absence of easily discernible changes in the slopes in $[f^*, \log r_f]$ plot it is difficult to justify such an analysis. Here, we show by analyzing experimental data that the observed non-linearity in the DFS data of several protein complexes can be better accounted for with an energy landscape containing multiple sequential barriers, as originally demonstrated \cite{EvansNature99,Moy06Biomacromolecules}.

\begin {figure}
 \includegraphics[width=3.50in]{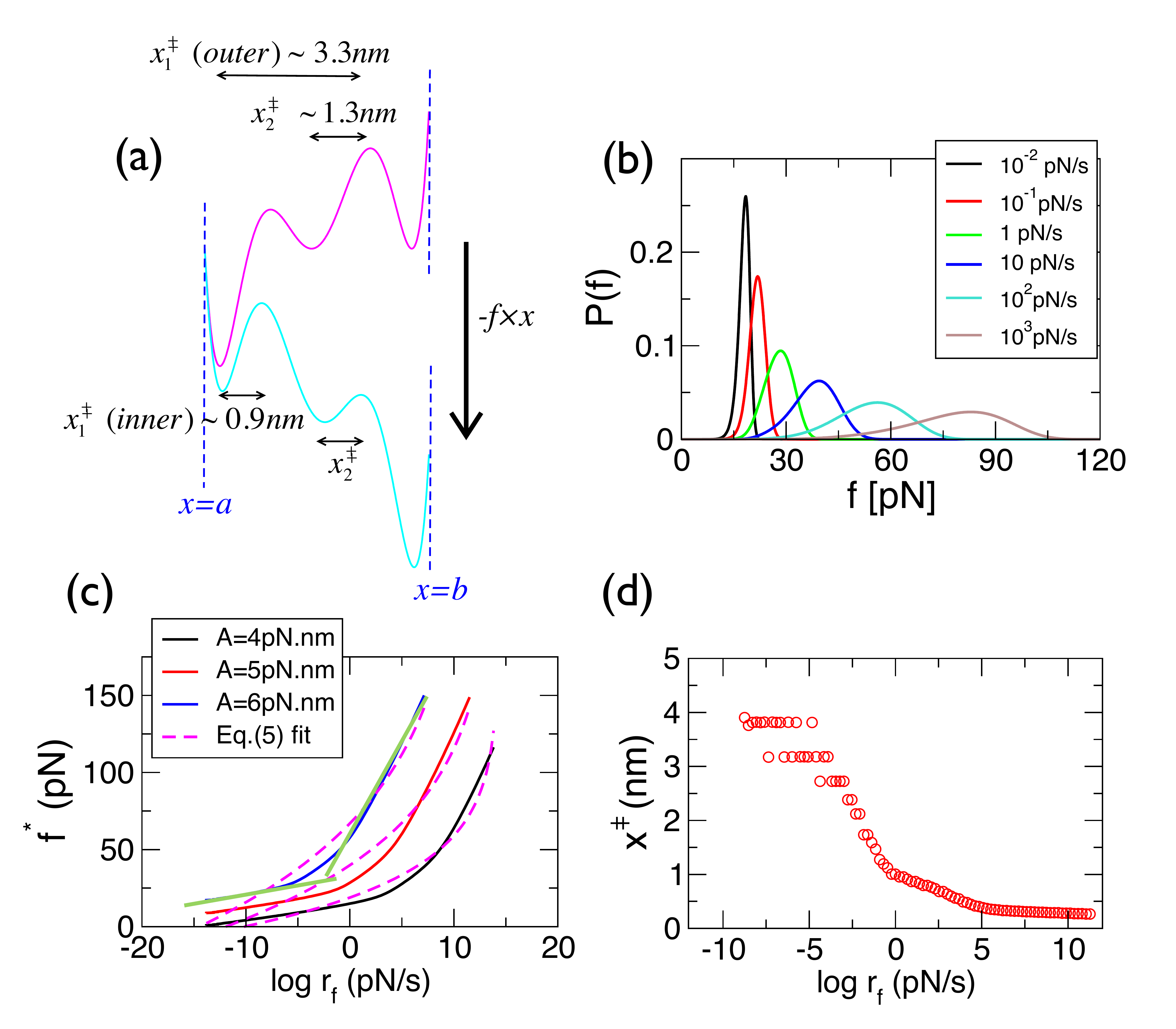}
 \caption{Rupture characteristics obtained numerically using a potential with two barriers at constant loading rates. (a) $U(x)$ (magenta) and $U(x)-f\cdot x$ (cyan) with $A = 5$ $pN\cdot nm$ and $f = 50$ pN. Reflecting and absorbing boundary conditions are set at $x=a$ and $x=b$, respectively. (b) Rupture force distributions, $P(f)=k(f)/r_f\cdot \exp{\left[-\int^f_0df'k(f')/r_f\right]}$, at varying $r_f$ were computed by using mean first passage time (MFPT), 
 $k^{-1}(f)=D^{-1}\int^b_adye^{\beta(U(y)-f\cdot y)}\int^y_adze^{-\beta(U(z)-f\cdot z)}$, starting from the first bound state at $a$(=0 nm) to reach an absorbing boundary at $b$(=5 nm). MFPT expression is valid in the force regime where stationary flux approximation holds \cite{Evans97BJ}. 
The length was scaled by nm, and $D=1.0\times 10^7$ $nm^2/s$ was used for the diffusion constant. (c) $[f^*,\log r_f]$ plots at three $A$ values. Fits of $[f^*,\log r_f]$ to Eq.\ref{eqn:f*} yield $\nu\ll 0.5$ for all $A$ values ($\nu=0.064$, $0.075$, $0.046$ for $A=4, 5, 6$ $pN\cdot nm$, respectively). 
In this case, the data should be divided into two regions and analyzed by the two linear fits as depicted using green lines on the curve with $A=6pN\cdot nm$. 
(d) loading rate dependent $x^{\ddagger}(r_f)(= x_{ts} -x_b)$, extracted from the slope of plot at each $r_f$ in (c) with $A=5$ $pN\cdot nm$, shows a sharp decrease from $\sim$3 nm to $<1$ nm around $r_f \approx (e^{-3} - e^0)$ pN/s.\label{fig:inverse}} 
\end{figure}


To illustrate how steep curvatures in DFS data can arise naturally from a 1D free energy profile we calculated $P(f)$ and $[f^*,\log r_f]$ of forced-escape kinetics of a quasiparticle from a potential with two barriers, $U(x)=Ax(x-1)[(x-2)(x-3)(x-4)(x-5)+1]$ with $A>0$ 
(Fig.\ref{fig:inverse}). The distributions $P(f)$ are typical of what is observed in experiments (Fig.\ref{fig:inverse}b). 
For all values of $A$, $[f^*, \log r_f]$ plots are curved although one could discern a modest change in slope (Fig.\ref{fig:inverse}c). The loading rate dependent $x^{\ddagger}(r_f)$, calculated from the slope of the data $k_BT/x^{\ddagger}(r_f)$ at each $r_f$ in Fig.\ref{fig:inverse}c, changes from $\sim$ 3 nm to $<1$ nm. The precipitous change in $x^{\ddagger}$ at $r_f\approx (e^{-3} - e^0)$ pN/s reflects the transition of the confining barrier from outer to inner barrier with an increasing force (see Fig.\ref{fig:inverse}a). In contrast, gradual change of $x^{\ddagger}$ in the range $0<\log r_f <10$ is most likely due to the movement of the inner transition state (see Fig.5C in Ref.(5)). 
Although it is straightforward to interpret that the two discrete slopes in Fig.\ref{fig:inverse}c (or the precipitous transition of $x^{\ddagger}$ in Fig.\ref{fig:inverse}d) are due to crossing two barriers since the underlying potential is given in Fig.\ref{fig:inverse}a, it is nontrivial to solve the inverse problem of unambiguously determining from 
$[f^*, \log r_f]$ plots and decide whether the underlying free energy profile has a single barrier with a moving transition state as $r_f$ increases or multiple barriers.

\begin {figure}
 \includegraphics[width=3.40in]{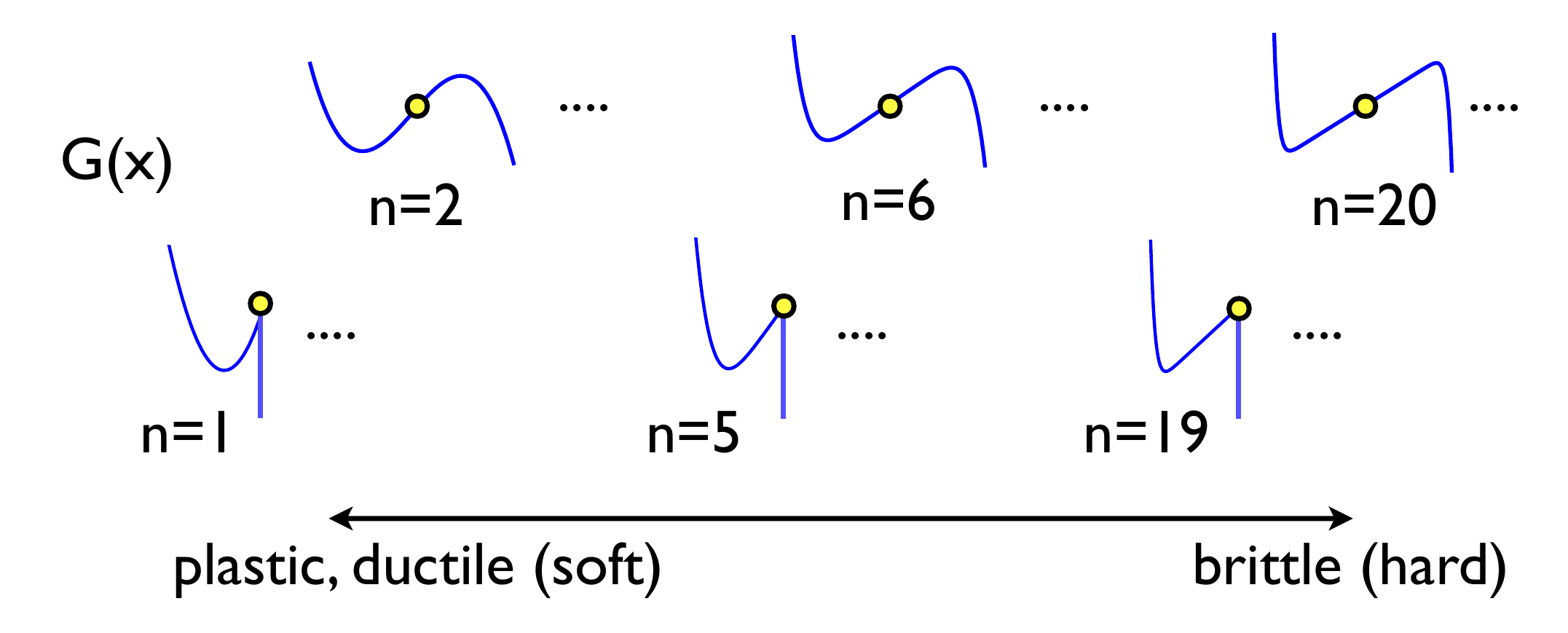}
 \caption{The $n$-dependent shape of $G(x)$ (Eq.\ref{eqn:Gpot}). The potential with increasing n is associated with more brittle molecular complexes. The yellow circle ($x = x_c$) denotes an inflection point and a cusp in each even and odd $n$ potential, respectively.
\label{fig:G_pot}
} 
\end{figure}

To establish a criterion for ascertaining whether the energy landscapes for forced-ligand rupture from biotin and LFA-1 have multiple barriers, we study the range of applicability  of DFS formalism based on a model potential with a single barrier. Consider a Kramers' problem of barrier crossing in a free energy profile $G(x)$ in which a single barrier separates the bound and unbound states of a quasi-particle as in  ligand bound in a pocket of a receptor: 
\begin{equation}
G(x)=G(x_c)+f_c(x-x_c)+\frac{(-1)^{n+1}M}{(n+1)!}(x-x_c)^{n+1}
\label{eqn:Gpot}
\end{equation}
with $M > 0$. 
In $G(x)$, a 1D free energy profile with a single barrier, the shape of barrier and energy well is approximated using $n$-th order polynomial with $n=1, 2, 3, \cdots$. 
For odd $n$, we assume that $G(x)=-\infty$ for $x>x_c$, so that the transition state of $G(x)$ is cusped. In the absence of tension, the barrier height, $G^{\ddagger}$, and the location of transition state, $x^{\ddagger}$, are $G^{\ddagger}=\chi \frac{n}{n+1} f_c(n!f_c/M)^{1/n}$ and $x^{\ddagger}=\chi(n! f_c /M)^{1/n}$, respectively, where $\chi=1$ (for odd n), 2 (for even n).  Thus $f_c=\frac{n+1}{n} G^{\ddagger}/x^{\ddagger}$. 
The form of $G(x)$, an extension of the microscopic theories using harmonic-cusp or linear-cubic potential, accounts for the degree of plasticity (or ductility) or brittleness of the energy landscape \cite{Evans97BJ} by changing $n$ (Fig.\ref{fig:G_pot}) \cite{Hyeon07JP}. 
Under tension, $G_{eff}(x)=G(x)-f\cdot x$; $f_c$ should be replaced with $f_c(1-f/f_c)=f_c\varepsilon$. Therefore, 
$G^{\ddagger}(f)=G^{\ddagger}\varepsilon^{1+1/n}$ and $x^{\ddagger}(f)=x^{\ddagger}\varepsilon^{1/n}$.  
Although Eq.\ref{eqn:Gpot} looks similar to the one Lin \emph{et al.} used to discuss rupture dynamics for $\varepsilon\ll 1$ where the barrier height is almost negligible \cite{Lin07PRL}, we did not impose any specific force condition on $G(x)$. Instead of attributing  the movement of transition state to a large external tension \cite{HummerBJ03,Dudko03PNAS,Sheng05JCP,Dudko06PRL,Lin07PRL,Friddle08PRL}, we mapped the non-linearity in DFS data onto $G(x)$ that has the $n$-dependent shape of transition barrier and bound state. In $G(x)$, increasing brittleness makes $x^{\ddagger}(f)$ insensitive to applied tension, which is dictated by $n$; $x^{\ddagger}(f)/x^{\ddagger}=\varepsilon^{1/n}\rightarrow 1$. 
For a generic free energy profile $F(x)$ with high curvatures at both $x=x_{ts}$ and $x_b$, 
$x^{\ddagger}(f)/x^{\ddagger}=1-f/x^{\ddagger}\times (|F^{\prime\prime}(x_{ts})|^{-1}+|F^{\prime\prime}(x_b)|^{-1})\rightarrow 1$ \cite{Hyeon07JP}. 
When free energy profile is associated with a brittle barrier, Bell's formula can be used to extract the feature of the underlying 1D profile from DFS data \cite{Hyeon07JP}. 

For general $n$, the KramersÕ rate equation based on the Eq.\ref{eqn:Gpot} under tension can be derived as: 
\begin{equation}
k(\varepsilon)=\kappa\varepsilon^{\alpha(n)}\exp{(-\beta G^{\ddagger}\varepsilon^{(n+1)/n})}
\end{equation}
where $\kappa$ is the prefactor in Kramers theory and $\alpha(n)=\chi(1-1/n)$ with $\chi=1$, 2 for odd and even $n$, respectively. 
For a given $k(f)$, the most probable unbinding force is determined by $dP(f)/df|_{f=f^*}=0$, resulting in a general equation for  $f^*$:
\begin{equation}
k'(f^*)=\frac{1}{r_f}[k(f^*)]^2
\end{equation}
which leads to 
\begin{align}
\varepsilon^{\frac{n+1}{n}}=\frac{-1}{\beta G^{\ddagger}}\log{\left[\frac{r_fx^{\ddagger}}{\kappa k_BT}\varepsilon^{1/n-\alpha(n)}\left(1-\frac{1}{\beta G^{\ddagger}}\frac{n\alpha(n)}{n+1}\varepsilon^{-\frac{n+1}{n}}\right)\right]}.
\end{align}
Under the typical condition that rupture occurs by thermal activation, i.e., $f\ll f_c (\varepsilon\approx 1)$ and $\beta G^{\ddagger}\gg 1$, the most probable unbinding force is approximated as:
\begin{equation}
f^*\approx f_c\left[1-\left(-\frac{k_BT}{G^{\ddagger}}\log\frac{r_fx^{\ddagger}}{\kappa k_BT}\right)^{\nu}\right]
\label{eqn:f*}
\end{equation}
where $\nu=\frac{n}{n+1}$. In deriving Eq.\ref{eqn:f*} using $G(x)$, the large force $\varepsilon(=1- f/f_c) \ll 1$ or fast loading condition, an assumption made in obtaining the mean unbinding force expression similar to Eq.\ref{eqn:f*} \cite{Garg95PRB,Lin07PRL,Friddle08PRL}, is not needed. 
Only the shape of the energy potential matters in deriving Eq.\ref{eqn:f*} from Eq.\ref{eqn:Gpot}.  
The DFS data will have a larger curvature for smaller $n$, namely when the energy landscape associated with a protein complex is more ductile (Fig. 2). Because $n=1$ (harmonic cusp), $2$ (linear cubic), $\ldots$, $\infty$ (Bell), $\nu$ must satisfy the bound, 
\begin{equation}
1/2\leq \nu \leq 1
\label{eqn:inequality}
\end{equation}
for an arbitrary 1D profile that suffices to describe rupture kinetics.

\begin {figure}
 \includegraphics[width=3.50in]{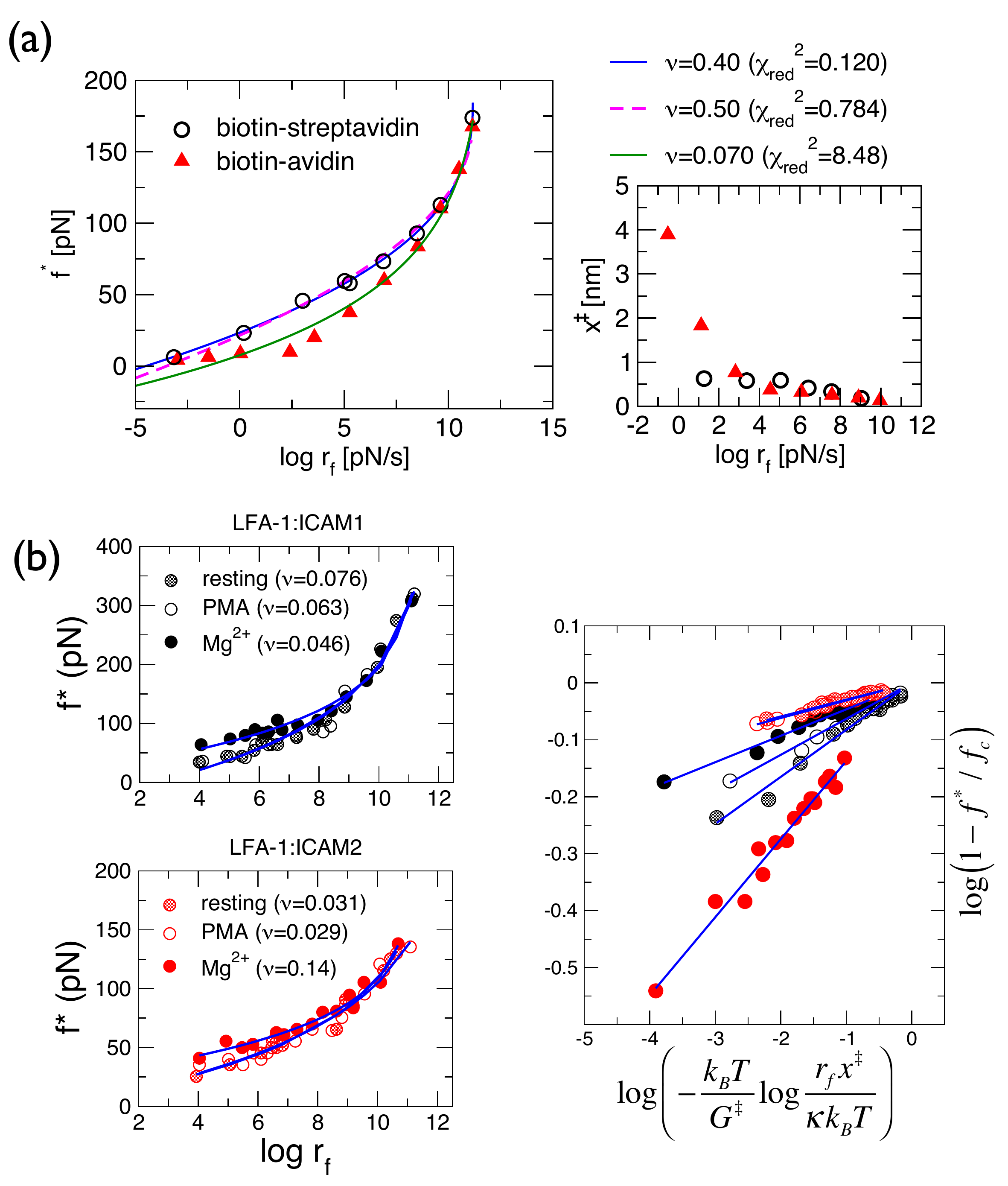}
 \caption{Analysis of DFS data with large curvatures. (a) The data obtained using biomembrane force probe (BFP) with force constant in the range 0.1-3 pN/nm \cite{EvansNature99} were fitted to Eq.\ref{eqn:f*} (solid lines) with $\nu$=0.40 for biotin-streptavidin (circle) and $\nu$=0.070 for biotin-avidin (triangle).   
The $x^{\ddagger}(r_f)$ at each $r_f$ is calculated on the right using the slope of four successive data points of $[f^*, \log r_f ]$ plot. 
Analyses of data using restricted $\nu$ values ($\nu=0.5$ fit is in dashed line in Fig.\ref{fig:Evans_Moy}a) are in the SI
(b) Analysis of DFS data of LFA-1 and its ligands, ICAM-1 and ICAM-2 in Ref. \cite{Moy06Biomacromolecules}. 
The fits in log-log scale are shown on the right.  
In all cases, $\nu<1/2$ suggests that for these complexes as well the underlying free energy profiles must contain at least two barriers; thus multi-state fits are required by dividing the DFS data into multiple regions as was already surmised in \cite{Moy06Biomacromolecules}. 
\label{fig:Evans_Moy}}
\end{figure}

For forced-rupture of biotin-ligand complex, fits to the entire range of the data using Eq.\ref{eqn:f*} give $\nu$ in the disallowed range; $\nu$ = 0.40 (biotin-streptavidin) and $\nu$ = 0.070 (biotin-avidin) (see Fig.\ref{fig:Evans_Moy}a). 
Even in biotin-streptavidin case, the parameters extracted from the fits with $\nu=0.40$ and $\nu=0.5$ (fixed) are comparable; however, the fit with $\nu=0.40$ is superior yielding both smaller relative error and reduced chi-square value, $\chi_{red}^2$, than with $\nu=0.5$, the lower bound of Eq.
\ref{eqn:inequality}, that gives the maximal curvature in the single-barrier picture (see Fig.\ref{fig:Evans_Moy}(a) and Fig.\ref{fig:compare} in the SI). 
For both biotin-ligand complexes, our criterion consistently suggests that the unbinding landscapes for the complexes involve more than one barrier, as was emphasized by Merkel \emph{et al.} \cite{EvansNature99}. 
Next, we analyzed the extensive data on LFA-1 expressed in Jurkat T cells whose binding affinity to ICAM-1 and ICAM-2 can be enhanced by treating the cells with phorbol myristate acetate (PMA) and the divalent counterion, Mg$^{2+}$. 
Under all conditions the exponents that best fit the  DFS data are $\nu<0.5$ (Fig.\ref{fig:Evans_Moy}b).  As originally argued by entirely different method \cite{Moy06Biomacromolecules} rupture of ICAM-1 and ICAM-2 from LFA-1 is best described using a free energy profile with at least two barriers. Taken together we arrive at a consistent conclusion that $\nu<0.5$ implies that the underlying free energy landscape in protein-ligand complexes has  multiple barriers.

Mathematically the inequality (Eq.\ref{eqn:inequality}) is not strict because it is possible to construct 1D profiles with $\nu < 0.5$ for which $[f^*,\log r_f]$ plots exhibit curvature like those observed in experiments. 
However, such free energy profiles are physically pathological with non-existing first derivatives in the vicinity of the bound complex and fits to the data give manifestly unrealistic parameters (see Supporting Information for detailed calculations and analysis). For the physical free energy profiles Eq.\ref{eqn:inequality} is rigorously satisfied. 
In addition, there is no compelling reason to choose a special $n$ value even if 1D profile is deemed adequate, and thus $\nu$ ought to be treated as a parameter. 
Although Ref. \cite{Dudko07BJ} used $\nu$ as a free parameter, the validity range for $\nu$ was not discussed. 
If a global fit of $[f^*, \log r_f]$ data using Eq.\ref{eqn:f*} yields $\nu< 0.5$ and the effect of probe stiffness \cite{Evans01ARBBS} is absent in the DFS data (see below), we can conclude that a single barrier description of the energy landscape is inadequate.

In principle curvature in the DFS data could also arise due to probe stiffness.  
Simple procedure of tiling free energy profile by the amount  $-f\cdot x$ under tension is widely used in analyzing single molecule force experiment.  
However, more rigorous formulation for the effective free energy profile under load using a transducer with stiffness $k$ should read $G_{tot}(x,X_{tr})=G(x)+\frac{1}{2}k_{eff}(x-X_{tr})^2$ where $x$ is the position of the end of molecule, $X_{tr}$ is the position of transducer, and $k_{eff}$ is the effective stiffness of molecular construct combining the transducer and the complex ($k_{eff}^{-1}=k_{tr}^{-1}+k_m^{-1}$).  In fact, $\frac{1}{2}k_{eff}(x-X_{tr})^2=-f\cdot x+\frac{1}{2}k_{eff}x^2+\frac{1}{2}k_{eff}X_{tr}^2$ with $f=k_{eff}X_{tr}$. 
Therefore, the effective free energy for the complex under tension should be written in general as  
$G_{eff}(x)=G(x)-[f-\frac{1}{2}k_{eff}x]\cdot x$ \cite{Walton08BJ}. 
As long as $f\gg \frac{1}{2}k_{eff}x$ (or $X_{tr}\gg x/2$) especially when $k_{eff}$ is small as in optical tweezers or BFP, one can approximate $G_{eff}(x)\approx G(x)-f\cdot x$.  
Otherwise, rebinding from transient capture well created by a large probe stiffness at near-equilibrium loading condition could give rise to the nonlinearity in the DFS data \cite{Evans01ARBBS}. 
Therefore, the rupture force being measured should be replaced by $f^*\rightarrow f^*-\frac{1}{2}k_{eff}x^{\ddagger}\varepsilon^{1/n}$, and at low forces ($f\ll f_c$) the most probable force measured by using a transducer with high stiffness such as AFM could be approximated as,
\begin{equation}
f^*\approx \underbrace{\frac{1}{2}k_{eff}x^{\ddagger}}_{=f_{pl}}+\underbrace{f_c\left[1-\left(-\frac{k_BT}{G^{\ddagger}}\log\frac{r_fx^{\ddagger}}{\kappa k_BT}\right)^{\nu}\right]}_{=f_{\mathrm{DFS}}}.
\end{equation}
The effect of probe stiffness manifests itself as a non-vanishing plateau force, which could be as large as $f_{pl}\approx (10-100)$ pN when $k_{eff}\approx 100$ pN/nm and $x^{\ddagger}=0.1-1$ nm, even when $r_f$ is small enough that $f_{\mathrm{DFS}}=0$ \cite{Evans01ARBBS}. 

The biotin-ligand complexes data preclude this possibility because the probe stiffness of BFP $k_{eff}=0.01-0.3$ pN/nm \cite{EvansNature99}, which is 1-2 orders of magnitude smaller than the $k_{eff}$ value discussed in the literature \cite{Friddle2008JPCC,Tshiprut2008BJ}. 
The value of $k_{eff}$ is $0.5-2.0$ pN/nm in the experiments involving LFA-1 \cite{Moy06Biomacromolecules}.  Even the largest estimated $x^{\ddagger}$ value for the outmost barrier ($x^{\ddagger}\approx 3$ nm) only yields $f_{pl}<1$ pN.   
Furthermore, if the nonlinear curvature of DFS data is suspected to be due to the stiffness effect, this ought to be discerned from the curvature due to multiple barriers by producing DFS data at a reduced probe stiffness. 
The curvature due to multiple barrier should persist in the DFS data even with a small $k_{eff}$. Thus, the curvature in the data in \cite{EvansNature99} can only be attributed to the presence of multiple barriers.


The condition (Eq.\ref{eqn:inequality}) for  single-barrier based 1D theories for DFS \cite{HummerBJ03,Dudko03PNAS,Sheng05JCP,Dudko06PRL,Dudko07BJ,Lin07PRL,Friddle08PRL} provides a guideline to judge whether the curvature in DFS data is due to multiple barriers or single barrier with a ductile transition state.  
Our work, which does not consider complications due to various multidimensional landscape scenarios \cite{Zhu03Nature,Barsegov05PNAS}, shows that the extracted parameters from the data for the protein complexes with ligands using 1D profile with multiple barriers are physically reasonable \cite{EvansNature99, Moy06Biomacromolecules}. Additional justification for the use of such energy landscapes can only be made by studying the structures of the protein complexes in detail.

\noindent {\bf Acknowledgements: }This work was funded by National Research Foundation of Korea (2010-0000602) (C.H.) and National Institutes of Health (GM089685) (D.T.). We thank Korea Institute for Advanced Study for providing computing resources.

\clearpage

\noindent{\bf SUPPORTING INFORMATION}\\

\noindent {\bf DFS theory for a free energy profile with non-integer $n$.}
It could be argued that the inequality $1/2\leq \nu\leq 1$ (valid rigorously for integer $n$)  that accounts for the curvature of DFS data is not mathematically required. Here we show that it is possible to construct 1D free energy profiles for which $\nu$ is clearly less than 0.5. Indeed, these free energy profiles can even have nearly vanishing $\nu$.  However, such profiles are unphysical because near the bound state they have  incorrect curvatures compared to the physical profiles discussed in the text and in the references cited therein. More importantly, the first derivatives  of these free energy profiles, which yield $\nu < 0.5$ do not exist near the bound state i.e, they have a singularity.  For these and other reasons (see below) we reject these free energy profiles as plausible models for explaining the curvatures in the observed [$f^*, \log{r_f}$] plots in a number of protein complexes discussed in the main text, which have all been explained using a two barrier model.

A free energy profile  with $0<n<1$ (see Eq.\ref{eqn:Gpot}) can lead to $0<\nu<1/2$ since $\nu=\frac{n}{n+1}$. 
To explore this possibility, we consider a free energy profile,
\begin{align}
G(x)=\sigma a x^{1/\sigma}-b x
\label{eqn:Gx}
\tag{S1}
\end{align}
with $\sigma>1$. Here we may regard $n=1/\sigma$, and hence $n < 1$. The term $-bx$ is required for the potential to have a barrier at a finite value of $x^{\ddagger}$, the location of the transition state (TS). 
The TS location  and the associated barrier height are 
\begin{align}
x^{\ddagger}&=\left(\frac{a}{b}\right)^{\frac{\sigma}{\sigma-1}}\nonumber\\
G^{\ddagger}&=
(\sigma-1)a\left(\frac{a}{b}\right)^{\frac{1}{\sigma-1}}.
\tag{S2}
\end{align}
Under tension $f$ the potential becomes $G_{eff}=G(x)-fx=\sigma a x^{1/\sigma}-(b+f)x$. 
The $f$-dependent TS location  and the barrier height are 
\begin{align}
\frac{x^{\ddagger}(f)}{x^{\ddagger}}&=\left(\frac{b}{b+f}\right)^{\frac{\sigma}{\sigma-1}}=\left(\frac{\frac{G^{\ddagger}/x^{\ddagger}}{\sigma-1}}{\frac{G^{\ddagger}/x^{\ddagger}}{\sigma-1}+f}\right)^{\frac{\sigma}{\sigma-1}}=\eta^{\frac{\sigma}{1-\sigma}}\nonumber\\
\frac{G^{\ddagger}(f)}{G^{\ddagger}}&=\left(\frac{b}{b+f}\right)^{\frac{1}{\sigma-1}}=\left(\frac{\frac{G^{\ddagger}/x^{\ddagger}}{\sigma-1}}{\frac{G^{\ddagger}/x^{\ddagger}}{\sigma-1}+f}\right)^{\frac{1}{\sigma-1}}=\eta^{\frac{1}{1-\sigma}}. 
\tag{S3}
\end{align}
where $\eta\equiv (1+f/f_{1/\sigma})$ with $f_{1/\sigma}\equiv \frac{1}{\sigma-1}\frac{G^{\ddagger}}{x^{\ddagger}}$. 
Therefore, one can rewrite Eq.\ref{eqn:Gx} ($G(x)$) and an effective free energy ($G_{eff}(x)$) under tension as  
\begin{align}
G(x)&=\frac{\sigma G^{\ddagger}}{\sigma-1}\left(\frac{x}{x^{\ddagger}}\right)^{1/\sigma}-\frac{G^{\ddagger}}{\sigma-1}\left(\frac{x}{x^{\ddagger}}\right),\nonumber\\
G_{eff}(x)&=\frac{\sigma G^{\ddagger}}{\sigma-1}\left(\frac{x}{x^{\ddagger}}\right)^{1/\sigma}-(\frac{G^{\ddagger}}{\sigma-1}+fx^{\ddagger})\left(\frac{x}{x^{\ddagger}}\right). 
\tag{S4}
\end{align}

Given $G_{eff}(x)$, it is possible to obtain the mean first passage time expression corresponding to the lifetime of the complex that can be measured in single molecule experiments. It  is given by, 
\begin{equation}
k(f)^{-1}=\frac{1}{D}\int^{\infty}_0dx e^{\beta G_{eff}(x)}\int^x_0dy e^{-\beta G_{eff}(y)}.
\tag{S5}
\end{equation}
The saddle-point approximation, 
$k(f)\approx D\left(\int_{bound}dy e^{-\beta G'_{eff}(0)y}\right)^{-1}\sqrt{\frac{G''_{eff}(x^{\ddagger}(f))}{2\pi k_BT}}e^{-\beta G^{\ddagger}(f)}$ with Eq.A3, yields Kramers' equation for the escape rate:  
\begin{align}
k(f)=\kappa \eta^{\alpha(\sigma)} \exp\left[{-\beta G^{\ddagger}\eta^{\frac{1}{1-\sigma}}}\right]
\tag{S6}
\end{align}
with $\alpha(\sigma)\equiv\frac{2\sigma-1}{2(\sigma-1)}$.
Here, note that the $\kappa$, defined as the prefactor, contains the singular integral $\left(\int_{bound}dy e^{-\beta G'_{eff}(0)y}\right)^{-1}$.
By using the relationship $k'(f^*)=[k(f^*)]^2/r_f$ to derive the most probable force, we get 
\begin{align}
\eta^{\frac{1}{1-\sigma}}=-\frac{1}{\beta G^{\ddagger}}\log{\frac{r_fx^{\ddagger}}{\kappa k_BT}\left[\frac{(\sigma-1)\alpha(\sigma)}{\eta^{\alpha(\sigma)+1}\beta G^{\ddagger}}+\eta^{\frac{\sigma}{1-\sigma}}\right]}.
\tag{S7}
\end{align}
By assuming $\beta G^{\ddagger}\gg 1$ and $f\ll f_{1/\sigma}$, we can simplify the above equation into
\begin{align}
\eta^{\frac{1}{1-\sigma}}\approx -\frac{1}{\beta G^{\ddagger}}\log{\frac{r_fx^{\ddagger}}{\kappa k_BT}}.
\tag{S8}
\end{align} 
Therefore, the most probable force ($f^*$) for the fractional potential (Eq.\ref{eqn:Gx}) is 
\begin{align}
f^*&\approx f_{1/\sigma}\left[\left(-\frac{1}{\beta G^{\ddagger}}\log{\frac{r_fx^{\ddagger}}{\kappa k_BT}}\right)^{1-\sigma}-1\right]
\nonumber\\
&=f_{1/\sigma}\left[\left(-\frac{1}{\beta G^{\ddagger}}\log{\frac{r_fx^{\ddagger}}{\kappa k_BT}}\right)^{2-1/\nu}-1\right], 
\label{eqn:Af*}
\tag{S9}
\end{align} 
where $\sigma=\frac{1-\nu}{\nu}$ was employed in the last line. 

There are a few important comments about Eq.\ref{eqn:Af*} that need to be made. (i) Note that the form of Eq.\ref{eqn:Af*} is very different from Eq.\ref{eqn:f*}. The difference is related to the aforementioned difficulties associated with $G_{eff}(x)$. Nevertheless, it is possible mathematically to construct model free energy profiles, without regard to physical considerations, for which [$f^*, \log{r_f}$] plot has curvature that is reminiscent of what is observed in experiments.  
(ii) Although Eq.\ref{eqn:Af*} can be used to obtain  excellent fits to  DFS data on protein complexes, it turns out that the extracted value of $\kappa$ is extremely large and are clearly unphysical (see Fig.\ref{fig:fraction}). The unphysical values are a consequence of the singularity of $G(x)$ (or $G_{eff}(x)$) at $x=0$. 
We conclude that the free energy profiles with a fractional power of $n$, which mathematically creates singularity at bound state, is not suitable to be used to analyze experimental data. Thus, the bound $1/2<\nu<1$ in Eq.\ref{eqn:inequality} must hold for physical 1D free energy profiles with a single barrier.

\onecolumngrid



\renewcommand{\thefigure}{S1}
 \begin {figure}
 \centering \includegraphics[width=6.0in]{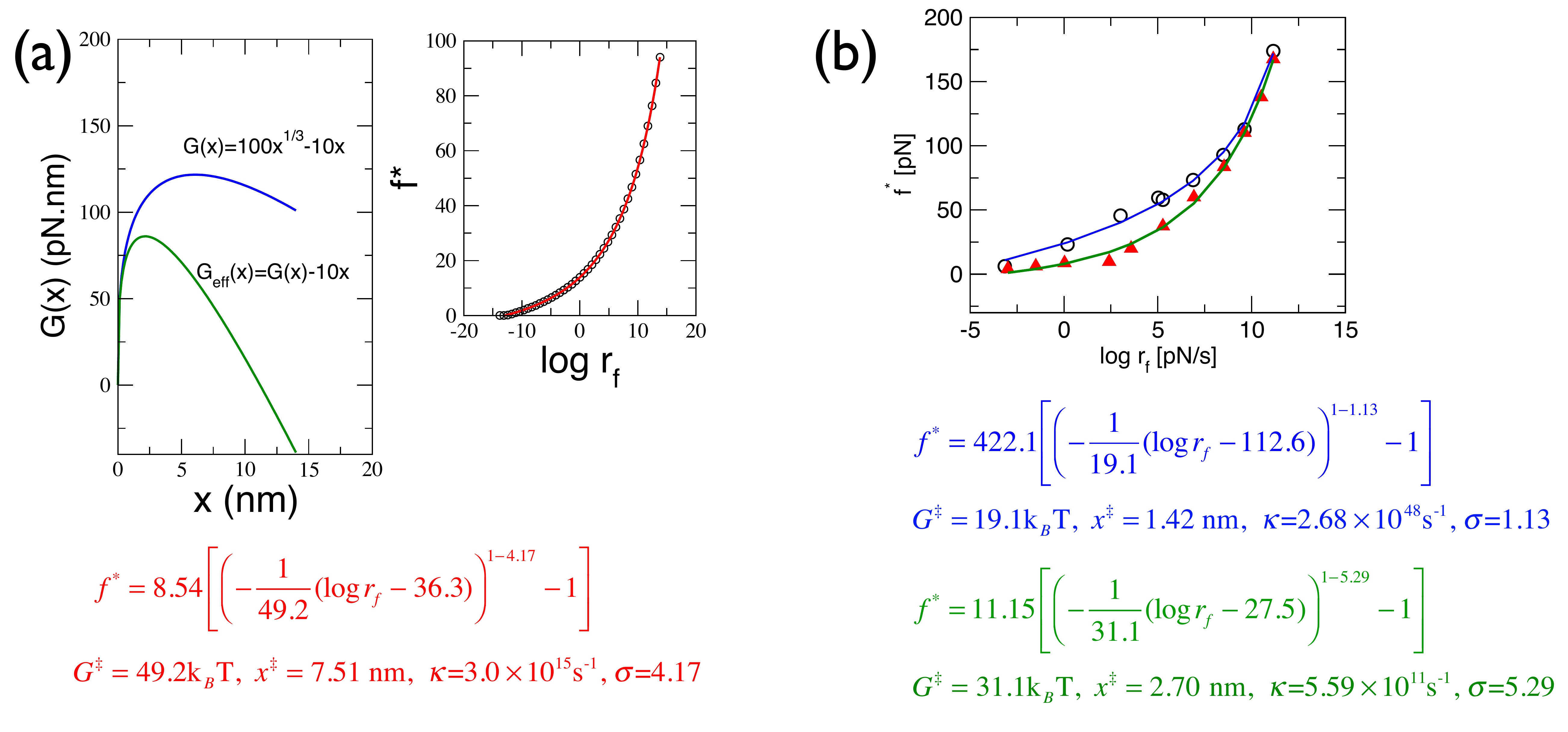}
 \caption{{\bf  (a)} The blue curve is the bare ($f$ = 0) free energy profile of the form given in Eq. (S1) and the green curve is the tilted form of $G(x)$ in the presence of force. By fitting the numerically computed (black circles) $f^*$ as a function of $r_f$ to Eq.\ref{eqn:f*} (red curve) we obtain the parameters shown below. Although the features of original potential $G(x)=100 x^{1/3}-10 x$ are reasonably recovered ($\sigma$ is larger than the value in $G(x)$)  by using Eq.\ref{eqn:f*}, the extracted value of $\kappa$ is unrealistically large. 
{\bf (b)} Eq.\ref{eqn:f*} was used to fit the DFS data of biotin-strepavidin (circles) and biotin-avidin (triangles). Although the quality of fit is excellent, the unrealistically large value of $\kappa$, due to the singularity of the hypothesized fractional potential at $x=0$, suggests that the potential with a fractional power should not be used for the analysis. In fact the $\kappa$ values are comparable to or much greater than the TST estimate $k_B T/h$ ($\approx 6.2\times 10^{12}$ $s^{-1}$), which of course makes no sense.  Hence,  we can rule out the free energy profiles of the form given in Eq.S1 to analyze DFS data on protein complexes.
\label{fig:fraction}} 
\end{figure}

\renewcommand{\thefigure}{S2}
 \begin {figure}[ht]
 \centering \includegraphics[width=6.5in]{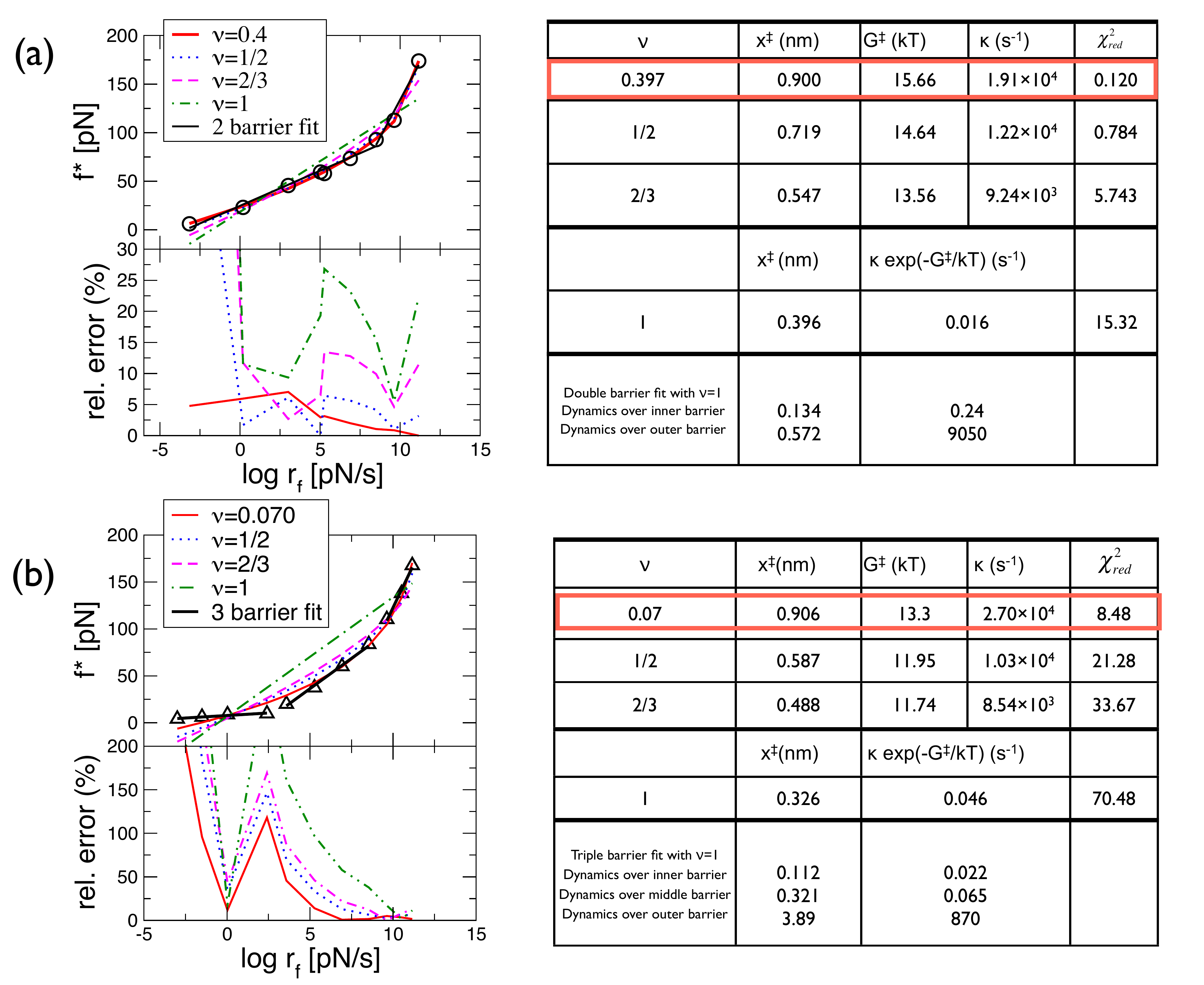}
 \caption{Analysis of DFS data using Eq.\ref{eqn:f*} for {\bf (a)} biotin-streptavidin and {\bf (b)} biotin-avidin. For each $\nu$, the fits, residuals ($|f^*_{fit}-f^*_{exp}|/f^*_{exp} \times100$), and extracted parameters were summarized in the table on the right. 
We can draw some general conclusions from the fits. For the biotin-streptavidin complex, fit with $\nu = 0.397$ produces the smallest errors although at high loading rates the relative errors for $\nu= 0.397$ and $\nu = 0.5$ are comparable. There are variations in other parameters ($x^{\ddagger}$, $G^{\ddagger}$, and $\kappa$) for all $\nu$. For the biotin-avidin complex the situation is far worse. In particular, the relative errors in the fits are large even when $\nu$ is varied. Similarly, the parameters extracted from the fits are not totally consistent.
Taken together, the fits using a 1D free energy profile with a single barrier is not appropriate to describe the rupture kinetics of these two complexes.  
\label{fig:compare}} 
\end{figure}

\end{document}